\documentclass[prl,twocolumn,showkeys,showpacs]{revtex4}
\usepackage{epsfig,graphicx,subfigure,url}%suthesis2,fancyhdr,a4}
\usepackage{amsmath,amsfonts}
\usepackage{mathptmx}
\pdfoutput=1
 \usepackage{url,color}
 \usepackage[pdftex,colorlinks]{hyperref}
 \definecolor{darkblue}{cmyk}{1,0,0,0.8}
 \definecolor{darkred}{cmyk}{0,1,0,0.7}
 \hypersetup{anchorcolor=black, citecolor=darkblue,
 filecolor=darkblue,
 menucolor=darkblue,pagecolor=darkblue,urlcolor=darkblue,linkcolor=darkblue} 
\usepackage{microtype}
\renewcommand{\epsilon}{\varepsilon}
\renewcommand{\epsilon}{e}
\newcommand{\R}{\mathbb{R}}
\newcommand{\hz}{\mathrm{Hz}}
\newcommand{\PD}{\operatorname{PD}}
\newcommand{\ms}{\mathrm{ms}}
\newcommand{\mm}{\mathrm{mm}}
\newcommand{\cm}{\mathrm{cm}}
\newcommand{\m}{\mathrm{m}}

\newcommand{\old}{\mathrm{old}}
\newcommand{\new}{\mathrm{new}}

\newcommand{\avg}{\operatorname{avg}}
\begin{document}
\title{Experimental continuation of periodic orbits through a fold}
\author{J. Sieber} \affiliation{School of Engineering, University of
Aberdeen, Kings College, Aberdeen, AB24 3UE, U.K.}
\author{A. Gonzalez-Buelga} \author{S.A. Neild} \author{D.J. Wagg}
\author{B. Krauskopf} \affiliation{Faculty of Engineering, University
of Bristol, Queen's Building, University Walk, Bristol, BS8 1TR, U.K.}

%%%%%%%%%%%%%%%%%%%%%%%%%%%%%%%%%%%%%%%%%%%%%%%%%%%%%%%%%%%%%
\begin{abstract}
  We present a continuation method that enables one to track or
  continue branches of periodic orbits directly in an experiment
  when a parameter is changed. A control-based setup in combination
  with Newton iterations ensures that the periodic orbit can be
  continued even when it is unstable. This is demonstrated with the
  continuation of initially stable rotations of a vertically
  forced pendulum experiment through a fold bifurcation to find the
  unstable part of the branch.
\end{abstract}

%%%%%%%%%%%%%%%%%%%%%%%%%%%%%%%%%%%%%%%%%%%%%%%%%%%%%%%%%%%%%

% generates the title
\maketitle

%%%%%%%%%%%%%%%%%%%%%%%%%%%%%%%%%%%%%%%%%%%%%%%%%%%%%%%%%%%%%

Characterizing a nonlinear dynamical system typically requires the
systematic investigation of stable and unstable steady-states and
periodic orbits in the relevant parameter region of the system.
When a mathematical model is available this task can be tackled
efficiently by performing a bifurcation analysis with the method of
numerical continuation. It allows one to find and follow (or continue)
solutions when varying a parameter --- a technique that can also be
used to map out stability boundaries (bifurcations) in multiple
parameters.  Several software packages are available for this task;
see the review papers \cite{D07,GK07} as an entry point to the
literature.

In physical experiments the use of continuation methods has proved
much more difficult.  One approach is a combination of system
identification and feedback control as applied by \cite{AWC94,SMK04}
to equilibria. In principle, it is also applicable to periodic orbits
\cite{OGY90} but, as is reported in \cite{WW00}, these methods do not
generally work well when applied to real physical experiments. An
alternative is extended time-delayed feedback (ETDF) \cite{P92,P01},
where the system is subject to a feedback loop with a delay that is
given by the period of the periodic orbit one wishes to stabilize.
This approach avoids system identification and, thus, is easier to
implement in real experiments \cite{SHWSH06}; see also the recent
collection of reviews \cite{SS07}.

An important prototype problem for experimental continuation is the
continuation of a stable periodic orbit through a fold (saddle-node
bifurcation). As one varies a system parameter the stable periodic
orbit gradually loses stability and then becomes unstable as it `turns
around' at the fold point. One problem is that ETDF and its modifications such
as described in \cite{P01} do not converge uniformly near a fold of
periodic orbits, meaning that they can generally not be used for tracking
 through a fold point; for a treatment
of the autonomous case see \cite{FFGHS07}.

We present and demonstrate here a continuation method that can be used
directly in an experiment to continue periodic orbits irrespective of
their stability.  Our method does not require a mathematical model nor
the setting of specific initial conditions. Instead it relies on
standard feedback control.  The feedback reference signal is updated
by a Newton iteration that converges to a state where the control
becomes zero.  The general ideas behind this method are described and
tested extensively in simulations in \cite{SK08b}. 

The implementation of feedback control requires one to measure some
output of the experiment with sufficient accuracy and to provide input
into the experiment in a tunable way. This requirement is quite
naturally satisfied, for example, for experiments in chemistry
\cite{SMK04,PCS94} and on electrical circuitry \cite{JBORB97}, as well
as for hybrid stability tests in engineering.  This type of test,
where a mechanical laboratory experiment of a critical component is
coupled bidirectionally to a numerical model of the remainder of the
tested system \cite{BWDW01}, is the motivating application behind the
development of experimental continuation methods \cite{SK08}.

%%%%%%%%%%%%%%%%%%%%%%%%%%%%%%%%%%%%%%%%%%%%%%%%%%%%%%%%%%%%%
\begin{figure}[t]
  \centering \includegraphics[width=0.85\columnwidth]{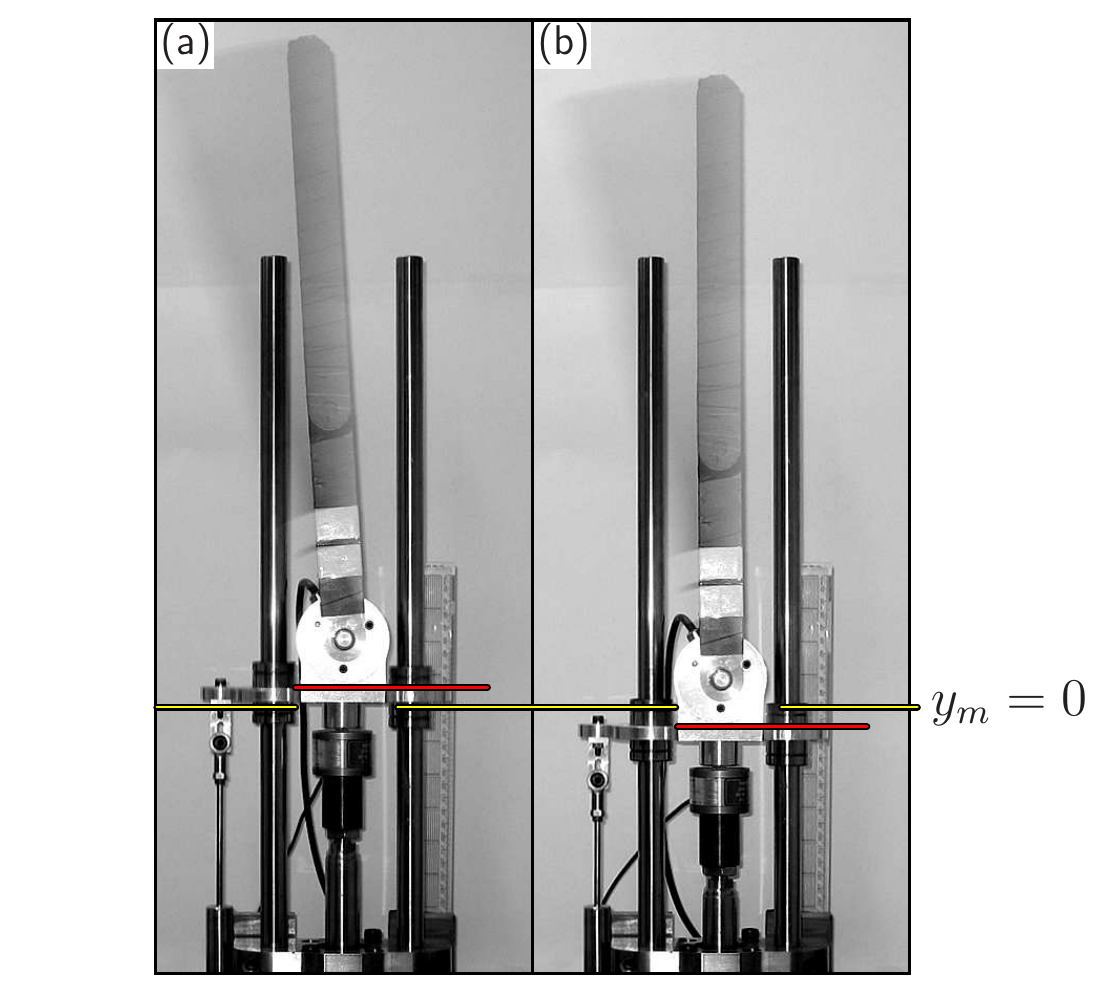}
  \caption{Photographs taken during continuation tests showing when
      the pendulum is at the top of a stable (a) and an unstable (b)
      rotation; the horizontal line ($y_m=0$) denotes the zero
      position.}
  \label{fig:photo}
\end{figure}
%%%%%%%%%%%%%%%%%%%%%%%%%%%%%%%%%%%%%%%%%%%%%%%%%%%%%%%%%%%%%

The goal of this paper is to demonstrate that our method can indeed be
used in an actual experiment to track periodic orbits reliably through
folds to reveal branches of unstable orbits. To this end, we
consider a classical mechanical experiment: the vertically forced
pendulum. 

In our experiment, a pendulum is attached to a pivot that moves
vertically along a trajectory $y_m(t)$, which is controlled via a
servo-mechanical actuator; this setup is as presented in \cite{GWN06}
and shown in the photos in figure~\ref{fig:photo}.  The actuator takes
a reference trajectory $y_r(t)$ as its input signal and aims to match
its output displacement $y_m(t)$ to this reference signal $y_r(t)$.
If
\begin{equation}\label{eq:unc}
  y_r(t)=p\sin(\omega t)
\end{equation}
then the pendulum is harmonically forced in the vertical direction
with forcing frequency $\omega$ and forcing amplitude $p$.  The
internal dynamics of the actuator translating the reference $y_r$ into
the actual motion $y_m$ is only known approximately.  However, when
$\omega$ is less than $10\,\hz$ and if the forces exerted by the
pendulum are small, the output $y_m$ closely follows $y_r$ with a
small time lag ($\approx20\,\ms$) and a small amplitude discrepancy
(less than $0.5\,\mm$). The dynamics of the angular displacement
$\phi$ of the pendulum are approximately a single-degree-of freedom
system.

We consider here the period-one {\em rotations} of the vertically
forced pendulum, which are periodic orbits where the pendulum goes
over the top once per forcing period. For any fixed forcing frequency
$\omega$ and sufficiently large value of the forcing amplitude $p$ one
finds a dynamically stable period-one rotation. A characteristic
feature of the stable rotations is the in-phase relationship between
the pendulum and the forcing: the pivot is up when the pendulum is in
the upside-down position; see Fig.~\ref{fig:photo}(a).  For the same
values of $\omega$ and $p$ one also finds an unstable rotation, which
is in anti-phase with the forcing; see Fig.~\ref{fig:photo}(b). Both
rotations are born (for a given, fixed $\omega$) in a fold bifurcation
at some specific value $p_f(\omega)$ of the forcing amplitude, where a
Floquet multiplier passes through $1$. Note that the fold point
$p_f(\omega)$ also depends on the damping; if the damping is small and
viscous then $p_f(\omega)\sim\omega^{-1}$ for large frequencies. (In
our experiment with a pendulum of approximate effective length
$0.28\,\m$ any frequency $\omega/(2\pi)\geq2\,$Hz is large in this
sense.)

In the experiment we measure $\phi$ and record the output
\begin{equation}
  \label{eq:phitotheta}
  \theta(t)=\phi(t)-\omega t\mbox{,}
\end{equation}
which is periodic for a periodic rotation (period one corresponds to a
period of $T=2\pi/\omega$).  The rotations are feedback stabilizable
by adding control to the actuator input $y_r$ in \eqref{eq:unc} based
on the difference between the measured relative angle $\theta(t)$ and
a periodic reference signal $\tilde\theta(t)$.  Note that feedback
control via $y_r$ cannot achieve global stabilization because the
amount of control is limited by the physical restriction of the
reference signal $y_r$ to amplitudes less than $3\,\cm$. However,
local feedback stabilization is sufficient for our purposes. Namely,
we superimpose the feedback on the harmonic forcing \eqref{eq:unc} by
setting the requested pivot trajectory $y_r$ to the solution of
\begin{equation}
  \label{eq:pd}
  \ddot y_r(t)=-\omega^2 p\sin(\omega t)+S(\phi(t))
  \PD[\theta-\tilde\theta](t)
\end{equation}
where $S(\phi)=1/\sin\phi$ if $|\sin\phi|>0.2$ and $0$ otherwise. The
factor $S$ ensures that control is only applied at non-zero rotation
angles ($\phi\neq0,\pi$).  The second term in \eqref{eq:pd} is a
standard proportional-plus-derivative (PD) controller defined by
$\PD[x]=k_px+k_d\dot x$ ($k_p=k_d=0.4$ in this experiment). Since the
angular velocity $\dot\phi$ is not directly measured, the term $\dot
x$ is approximated by a linear filter $x_v=N\cdot(x-x_f)$ where $x_f$
is the solution of
\begin{math}
   \label{eq:df} \dot x_f=N\cdot(x-x_f)
\end{math}
and $N$ is a large quantity ($N=100$ in this experiment).  Equation
\eqref{eq:pd} and the filter are linear and are solved in real-time in
parallel with the experiment on a dSpace\,DS1104\,RD real-time
controller board.  To ensure that the solution of \eqref{eq:pd} meets
the physical restrictions on the actuator amplitude ($y_m\leq3\,\cm$)
we reset $\dot y_r$ whenever $\phi=0$.

The introduction of feedback control into the experiment via
\eqref{eq:pd} adds a parameter to the overall system: the (periodic)
reference signal $\tilde\theta(t)$.  We introduce the scalar parameter
$\tilde\theta_0$ and determine $\tilde\theta(t)$ using the recursion
relation (also evaluated in real time)
\begin{align}
    \tilde \theta_h(t)&=(1-R)\tilde\theta_h(t-T)+
    R\,[\theta(t-T)-\avg[\theta](t-T)]\nonumber\\
    \tilde\theta(t)&=\tilde\theta_0+\tilde\theta_h(t)  \label{eq:tdasp}
\end{align}
where $T=2\pi/\omega$ is the period of the forcing, $R\in(0,1]$ is a
relaxation factor and
\begin{math}
 % \label{eq:avg}
  \avg[\theta](t)=1/T\int_{t-T}^t\theta(\tau)\,\mathrm{d}\tau
\end{math}
is the average of the output $\theta$ over the last forcing period (it
is a constant scalar for $T$-periodic functions).  We define the limit
\begin{equation}
  \label{eq:Thetadef}
  \Theta(p,\tilde\theta_0):=\lim_{t\to\infty}\avg[\theta](t),
\end{equation}
which exists (and the convergence of the time profile is uniform) for
all pairs $(p,\tilde\theta_0)$ that are in the vicinity of the
(unknown) family of rotations near fold points.  Choosing $R$ closer
to zero enlarges the region where the limit \eqref{eq:Thetadef} exists
but slows down the convergence.  

Equation~\eqref{eq:Thetadef} defines a smooth map
$\Theta:\R^2\mapsto\R$ that maps the system parameter pair
$(p,\tilde\theta_0)$ to the asymptotic average of the output of the
experiment. The map $\Theta$ is not known analytically but can be
evaluated for any $(p,\tilde\theta_0)$ by running the experiment with
control \eqref{eq:pd} and \eqref{eq:tdasp} until the transients have
died out.  In practice the limit $\Theta(p,\tilde\theta_0)$ is reached
after 2--3 seconds during our experimental runs.

The reference signal $\tilde\theta(t)$ corresponds to a natural
periodic rotation of the original (uncontrolled) vertically forced
pendulum if and only if the difference $\theta-\tilde\theta$ is zero,
making the feedback control non-invasive. This is the case when the
fixed point equation
\begin{equation}
  \label{eq:fixedpoint}
  \Theta(p,\tilde\theta_0)-\tilde\theta_0=0
\end{equation}
is satisfied. For parameter pairs $(p,\tilde\theta_0)$ satisfying
  \eqref{eq:fixedpoint} the parameter $\tilde\theta_0$ is equal to the
  average of the phase difference between the rotation and the
  forcing.

Our scheme is a modification of the classical ETDF scheme
\cite{P92,GSCS94}. The core of this modification is the solution of
the fixed point problem \eqref{eq:fixedpoint} by means of a Newton
iteration.  Classical ETDF corresponds for small $R$ and a fixed $p$
to a relaxed fixed point iteration
$\tilde\theta_{0,\new}=(1-R)\tilde\theta_{0,\old}+R
\Theta(p,\tilde\theta_{0,\old})$ for equation~\eqref{eq:fixedpoint},
which is known to diverge for the unstable rotations \cite{SS07}.  At
the fold point $(p_f,\tilde\theta_{0,f})$ the partial derivative
$\partial_2\Theta$ equals 1, and this makes the fixed-point problem
\eqref{eq:fixedpoint} singular.

To overcome this singularity we embed \eqref{eq:fixedpoint} into a
pseudo-arclength continuation \cite{D07}. The pairs
of $(p,\tilde\theta_0)$ satisfying \eqref{eq:fixedpoint} form a curve. We
introduce $y=(p,\tilde\theta_0)^T$, and extend \eqref{eq:fixedpoint}
by the \emph{pseudo-arclength condition}
\begin{equation}
  \label{eq:arc}
  y_t^T(y-y_\old)=h
\end{equation}
where $h$ is the (small) stepsize along the curve, $y_\old$ is the
previous point along the curve and $y_t$ is the unit secant through
the previous two points along the curve (as a practical approximation
of the tangent to the curve). Equations~\eqref{eq:fixedpoint}
and~\eqref{eq:arc} define a system of equations of the form $F(y)=0$,
which is uniformly regular near the fold. It can be solved by a
relaxed quasi-Newton recursion and we choose recursion with Broyden's
rank-one update; see \cite{SK08b}.

To start a continuation we choose a large forcing amplitude $p$
($2\,\cm$). Then the stable rotation of the uncontrolled system can be
found by swinging up the pendulum manually.  We measure the periodic
output $\theta$ and set the initial parameter $\tilde \theta_0$ to the
average of this output, thus defining the initial
$y=(p,\tilde\theta_0)^T$. In the actual implementation we scale $p$ by
a factor of $20$ so that both components of the vector $y$ are of
order one; the approximate initial secant to the curve is set to
$y_t=(-1,0)^T$. % The initial guess for the quasi-Newton Jacobian is
% $J=\left[
%   \begin{smallmatrix}
%     -1&0\\ \phantom{-}0&1
%   \end{smallmatrix}
% \right]$. 

%%%%%%%%%%%%%%%%%%%%%%%%%%%%%%%%%%%%%%%%%%%%%%%%%%%%%%%%%%%%%
\begin{figure}[t]
  \centering
  \includegraphics[width=0.65\columnwidth]{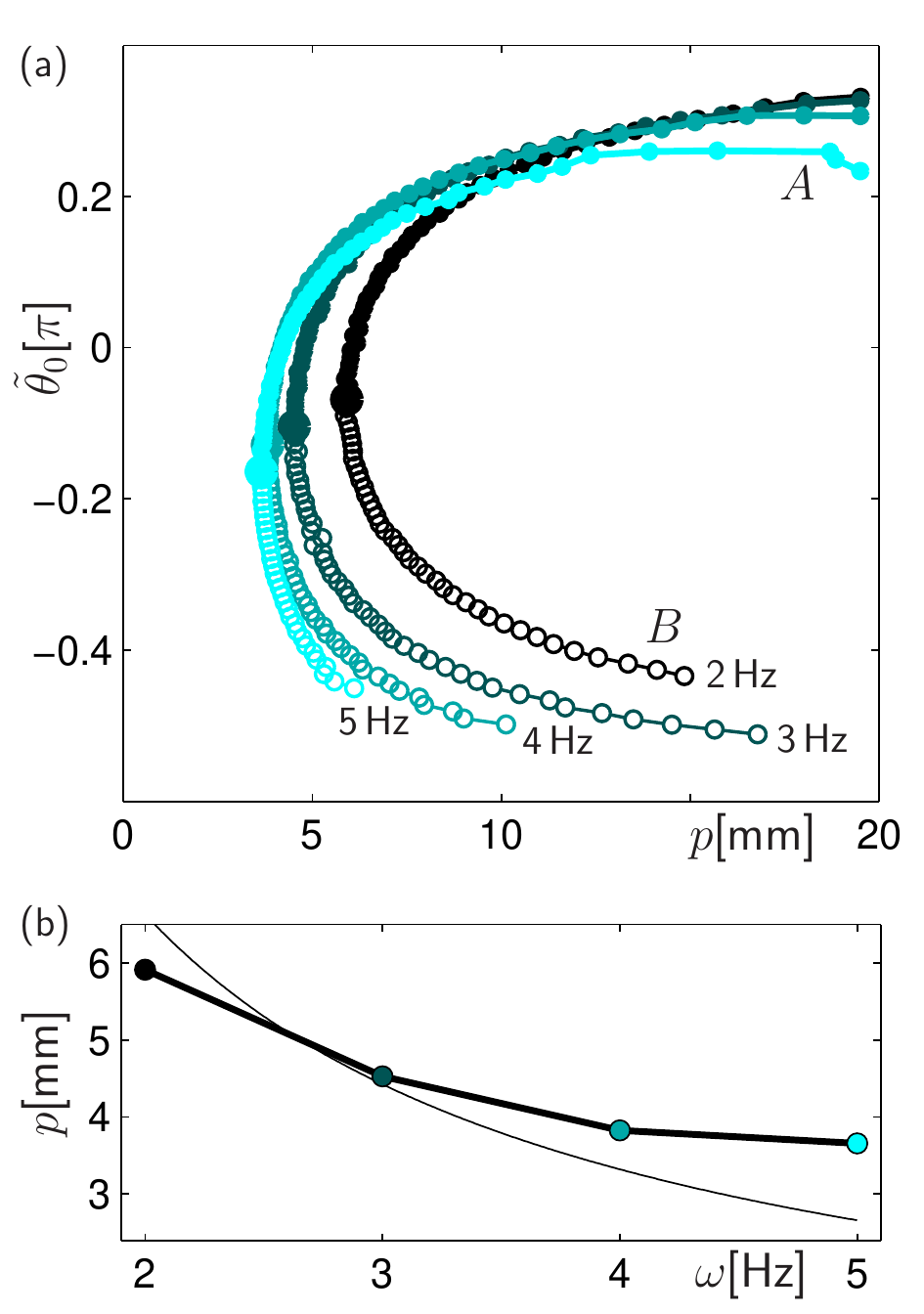}
  \caption{Experimental one-parameter bifurcation diagrams (a) for
    2\,Hz, 3\,Hz, 4\,Hz, 5\,Hz, respectively, showing 
    measured rotations (small circles: hollow for saddle rotations,
    full for stable rotations) and estimated fold points (large full
    circles). Panel (b) shows the fold points in $(\omega,p)$-plane
    (circles) and a viscous model estimate (thin solid
    line). Parameters values in
    \eqref{eq:pd},\,\eqref{eq:tdasp},\,\eqref{eq:arc} were
    $k_p=k_d=0.4$, $R=0.8$, $h=0.02$, and convergence tolerance
    $5\times10^{-3}$.}
  \label{fig:ampphas}
\end{figure}
%%%%%%%%%%%%%%%%%%%%%%%%%%%%%%%%%%%%%%%%%%%%%%%%%%%%%%%%%%%%%

Figure~\ref{fig:ampphas}(a) shows four branches of rotations in the
$(p,\tilde\theta_0)$-plane as continued by our method. Each branch is
for a different, fixed forcing frequency $\omega$ and varying forcing
amplitude $p$, continued from a stable rotation near the point $A$
through the fold to an unstable rotation near the point $B$. The upper
part of a branch corresponds to stable and the lower part to unstable
rotations. The larger circles on each of the branches in panel (a) are
the approximate values of the fold points $p_f(\omega)$.
Figure~\ref{fig:ampphas}(b) shows the location of the fold points in
the $(\omega,p)$-plane in comparison with the theoretical prediction
(thin solid curve) based on a viscous damping approximation.

Each of the four branches in Fig.~\ref{fig:ampphas}(a) is made up of
points at which the quasi-Newton recursion has converged; in practice
we accept a point when the difference $\avg[\theta]-\tilde\theta_0$
(which is the residual of equation \eqref{eq:fixedpoint}) stays below
$5\times10^{-3}$ during one forcing period. A continuation run is
performed as one continuous experiment without stopping or manual
intervention; it takes about 20 minutes for a curve resolution as in
Fig.~\ref{fig:ampphas}(a). The experimental continuation stops at the
lower end point of the branches, where the recursion \eqref{eq:tdasp}
becomes unstable at a period doubling.  This is a similar effect as
for the classical ETDF recursion, which has been found to lose
stability in a torus bifurcation \cite{JBORB97}.

%%%%%%%%%%%%%%%%%%%%%%%%%%%%%%%%%%%%%%%%%%%%%%%%%%%%%%%%%%%%%
\begin{figure}[t]
  \centering
  \includegraphics[width=0.55\columnwidth]{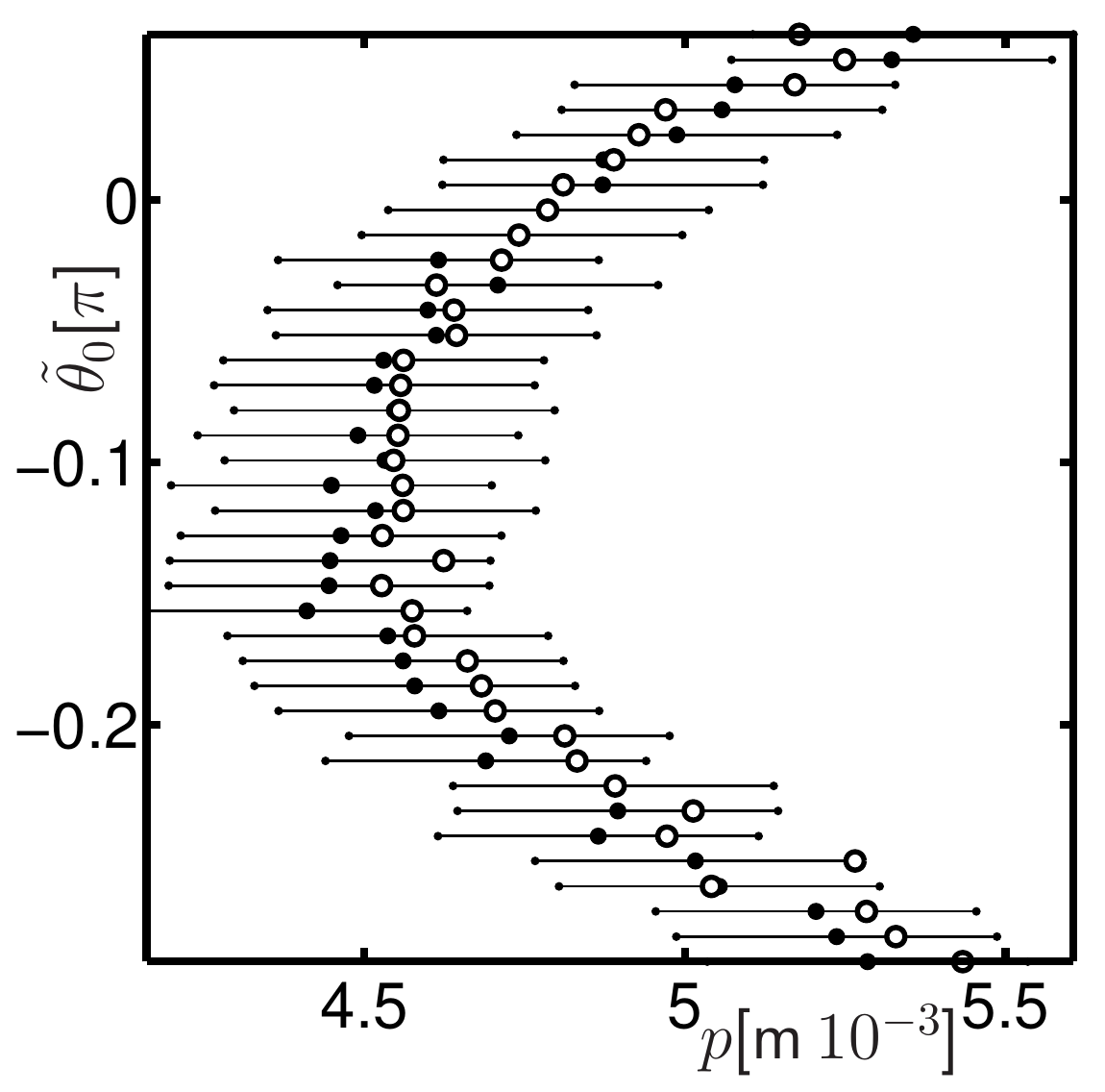}
  \caption{Variation of the phase compared to experimental accuracy
      near the fold for $\omega=3\,$\,Hz. The error bars indicate the
      maximum of $|p-p_m|$, where $p_m$ is the amplitude of the pivot
      displacement $y_m$.  Hollow circles: parameter $p$ as obtained
      by quasi-Newton iteration; full circles: $p_m$ as measured.}
  \label{fig:errbar}
\end{figure}
%%%%%%%%%%%%%%%%%%%%%%%%%%%%%%%%%%%%%%%%%%%%%%%%%%%%%%%%%%%%%

Figure~\ref{fig:errbar} shows an enlargement of the branch near the
fold for a forcing frequency of $\omega=3\,$\,Hz. Horizontal error
bars have been attached to each point (the vertical error in
$\tilde\theta_0$ is invisibly small). Their size highlights the
extreme difference in the scale of the axes: the range of $p$ is
$1$\,mm, which is of the order of a few multiples of the experimental
accuracy, whereas $\tilde\theta_0$ spans a range of approximately $60$
degrees.  This implies that in a small parameter region of $p$ near
the fold, between $4.5\,\mm$ and $5.5\,\mm$, the average phase
$\avg[\theta]$ of the rotation relative to the forcing changes by $60$
degrees.  Thus, the fold scenario presented in Fig.~\ref{fig:errbar}
is an example of a very sensitive dependence of the response (the
phase of the rotation) of a nonlinear dynamical system on its system
parameter (the forcing amplitude $p$). This implies that the rotations
shown in Fig.~\ref{fig:errbar} would be extremely difficult to find by
careful parameter tuning with the available experimental equipment
even on the stable part of the branch near the fold. By contrast, our
continuation method follows the branch of rotations through the rapid
change without difficulty: the dependence of the feedback controlled
pendulum on the parameter pair $(p,\tilde\theta_0)$ is not sensitive
and the resulting nonlinear system
\eqref{eq:fixedpoint}--\eqref{eq:arc} is uniformly well-conditioned
near the fold.

%%%%%%%%%%%%%%%%%%%%%%%%%%%%%%%%%%%%%%%%%%%%%%%%%%%%%%%%%%%%%
\begin{figure}[t]
  \begin{center}
    {\includegraphics[width=0.85\columnwidth]{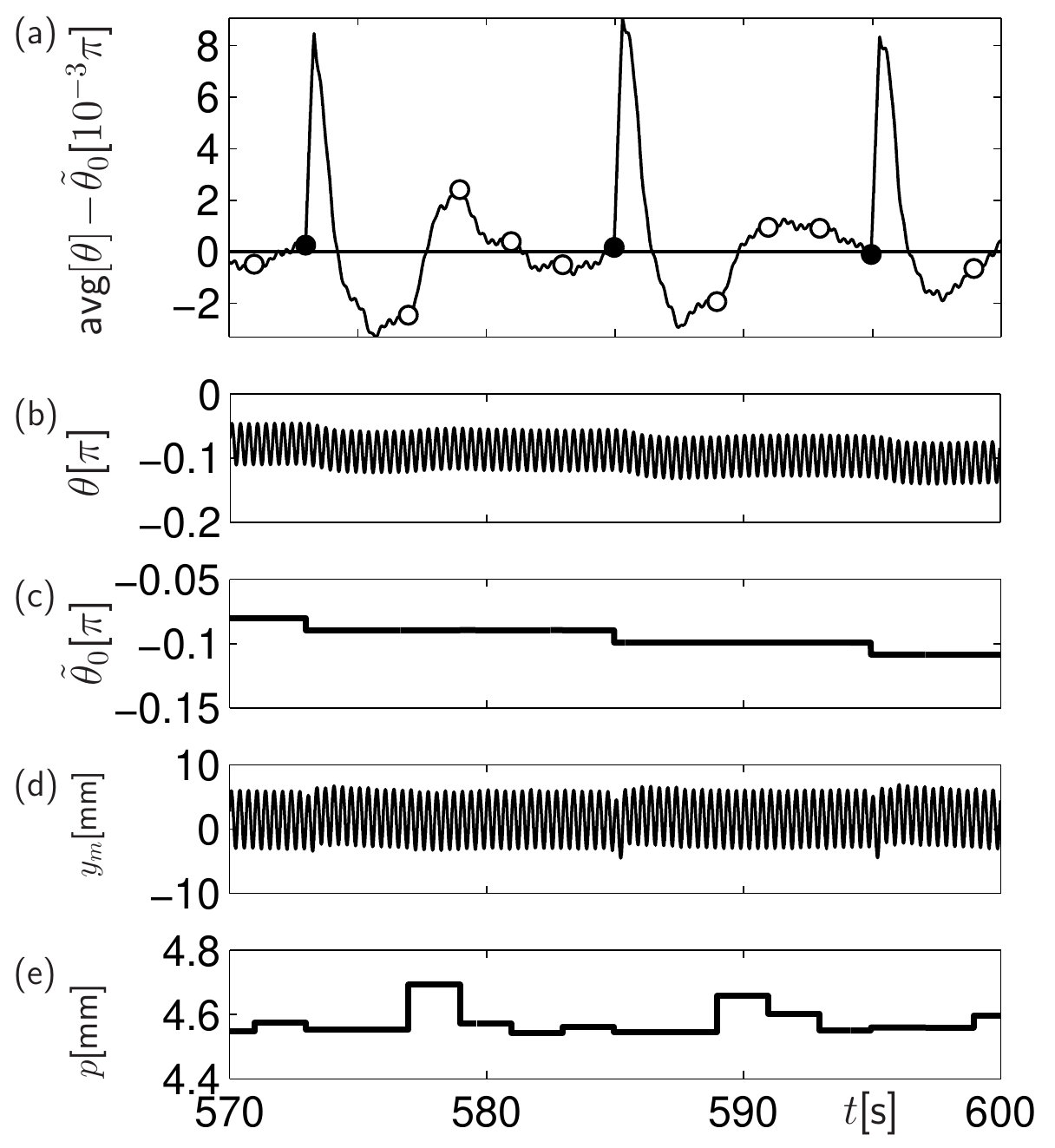}}
  \end{center}
  \caption{Time profiles during continuation for $\omega/(2\pi)=3$\,Hz
   [(a,\,b,\,d) measured, (c,\,e) set by quasi-Newton iteration].}
  \label{fig:timeprof}
\end{figure}
%%%%%%%%%%%%%%%%%%%%%%%%%%%%%%%%%%%%%%%%%%%%%%%%%%%%%%%%%%%%%

To provide more insight into how points on branches are accepted,
Fig.~\ref{fig:timeprof} shows a $30$\,s snapshot of the time profile
of the experimental continuation run for $3\,$\,Hz.  Panel~(a) shows
the measured difference $\avg[\theta]-\tilde\theta_0$, panel~(b) the
output $\theta$, panel~(d) the measured motion $y_m$ of the pivot, and
panels (c) and (e) the quantities $\tilde\theta_0$ and $p$ as updated
by the quasi-Newton iteration at discrete times.  Filled circles in
Fig.~\ref{fig:timeprof}(a) indicate when the difference
$\avg[\theta]-\tilde\theta_0$ is accepted as sufficiently small.  Then
the respective point $(p,\tilde\theta_0)$ is accepted and we start the
next step along the branch (by updating $y_\old$ and $y_t$ in the
pseudo-arclength condition \eqref{eq:arc}).  As a result, the
difference $\avg[\theta]-\tilde\theta_0$ jumps briefly to a much
larger value. The Newton iteration then drives the system to
convergence; the open circles indicate when $\theta-\tilde\theta$ has
been accepted as periodic. At these points $\avg[\theta]$ is measured
and new parameters $p$ and $\tilde\theta_0$ are set to initiate the
next Newton iterate.

In conclusion, we have presented a control-based continuation method
and demonstrated that it is capable of tracking periodic orbits
through fold bifurcations in a vertically forced pendulum experiment.
Our approach does not require knowledge of an underlying mathematical
model. Instead, we measure the amount of control and apply a Newton
iteration to drive the control action to zero to find the next point
on a branch.  Importantly, this Newton iteration does not have to run
in real-time, so that our method can be applied to any experiment that
is feedback stabilizable. Our ongoing work focuses on control-based
continuation of solutions and bifurcations in mechanical hybrid tests.
It would be an interesting challenge to investigate how our approach
could be extended to other application areas, such as neuroscience or
cell biology, where feedback control is generally more difficult to
achieve.

%\bibliography{delay}

\end{document}